\documentclass[11pt]{article}
\usepackage{mymoriond,epsfig}

\def\be{\begin{equation}}
\def\ee{\end{equation}}
\def\bea{\begin{eqnarray}}
\def\eea{\end{eqnarray}}

\newcommand{\FF}{{\cal F}_{\pi^0\gamma^*\gamma^*}}
\newcommand{\lag}{{\cal L}}
\newcommand{\order}{{\cal O}}
\newcommand{\lapprox}{%
\mathrel{%
\setbox0=\hbox{$<$}
\raise0.6ex\copy0\kern-\wd0
\lower0.65ex\hbox{$\sim$}
}}

\begin{document}
\vspace*{2cm}
\title{
     \begin{flushright}
       \textnormal{May 2003} 
     \end{flushright}  
    \vspace*{1.2cm}
    THEORETICAL STATUS OF THE MUON $g-2$~\footnote{Talk given at the 38th
    Rencontres de Moriond on Electroweak Interactions and Unified Theories, Les
    Arcs, France, 15-22 March 2003.}}

\author{ A. NYFFELER }

\address{Institute for Theoretical Physics, ETH Z\"urich \\
CH-8093 Z\"urich, Switzerland \\
nyffeler@itp.phys.ethz.ch}

\maketitle\abstracts{We review the present status of the theoretical
  evaluation of the anomalous magnetic moment of the the muon within the
  Standard Model. We mainly focus on the hadronic contributions in the muon
  $g-2$ due to vacuum polarization effects, light-by-light scattering and
  higher order electroweak corrections. We discuss some recent calculations
  together with their uncertainties and limitations and point out possible
  improvements in the future. In view of the inconsistent values for the
  hadronic vacuum polarization based on $e^+ e^-$ and $\tau$ data, no
  conclusion can be drawn yet, whether the apparent discrepancy between the
  current experimental and theoretical values for the muon $g-2$ points to
  physics beyond the Standard Model.}


\section{Introduction}
\label{sec:intro}

For a particle with spin $1/2$, the relation between its magnetic moment and
its spin reads $\vec \mu = g (e/2m) \vec s$. The Dirac equation predicts for
the gyromagnetic factor $g=2$, but radiative corrections to the
lepton-photon-lepton vertex in quantum field theory can shift the value
slightly. The anomalous magnetic moment is then defined as $a \equiv (g-2)/2$.
There has been a fruitful interplay between experiment and theory over many
decades. The results for the anomalous magnetic moments of leptons have
provided important insights into the structure of the fundamental interactions
(Dirac equation, QED, Standard Model, \ldots).

As we will briefly discuss below, the electron anomalous magnetic moment $a_e$
provides a stringent test of QED and leads to the most precise determination
of the fine structure constant $\alpha$. A weighted average~\cite{Mohr_Taylor}
of various measurements for $a_{e^+}$ and $a_{e^-}$, which is dominated by the
latest results of Ref.~\cite{a_e_measurements}, leads to
\be \label{a_e_exp}
a_e^{\mbox{{\scriptsize exp}}} = 11~596~521~88.3(4.2) \times 10^{-12}
\qquad \mbox{[3.7~ppb]} .   
\ee

The anomalous magnetic moment of the muon $a_\mu$, on the other hand, allows
to test the Standard Model as a whole, since all sectors contribute. The
current experimental world average, dominated by the recent measurements of
the $g-2$ collaboration at Brookhaven, reads~\cite{Gray}
\be \label{a_mu_exp} 
a_\mu^{\mbox{{\scriptsize exp}}} = 11~659~203(8) \times 10^{-10} 
\qquad \mbox{[0.7~ppm]} .   
\ee
The final goal is to reach an experimental precision of $4 \times 10^{-10}$.
In principle, $a_\mu$ is very sensitive to new physics beyond the Standard
Model. Since $a_l$ is dimensionless, one expects in general $a_l \sim (m_l /
M_{{\rm NP}})^2$, therefore $a_\mu$ is about $(m_\mu/m_e)^2 \sim 4\times 10^4$
times more sensitive to the scale of new physics, $M_{{\rm NP}}$, than $a_e$,
which makes up for the factor of 200 due to less precision in the measurement.
Unfortunately, the hadronic contributions lead to the largest source of error
in the Standard Model prediction for $a_\mu$, about $8 \times 10^{-10}$, and
they are very difficult to control. This hinders at present all efforts to
extract a clear sign of new physics from $a_\mu$.  The hadronic contributions
will be the main topic of this article. For more details on the subject of
$g-2$, we refer to the recent
reviews~\cite{review_C_M,review_Knecht,recent_reviews2}, whereas the
developments before 1990 can be traced from Ref.~\cite{older_reviews}. We will
not discuss at all here potential new physics contribution to $a_\mu$, like
supersymmetry, but refer instead to the article~\cite{New_Physics} and
references therein.


\section{Electron anomalous magnetic moment and $\alpha$}

The result for $a_e$ with up to four loops in QED can be written as 
follows 
\bea
a_e & = & {1\over 2} \left( {\alpha \over \pi} \right)
- 0.32847844400\ldots \left( {\alpha \over \pi} \right)^2 
+ 1.1812340\ldots \left( {\alpha \over \pi} \right)^3 
- 1.7502(384)\left( {\alpha \over \pi} \right)^4  \nonumber \\
&& + 1.70(3) \times 10^{-12} \, .  \label{a_e_th}
\eea
Above we have also included small, mass-dependent corrections due to internal
vacuum polarization diagrams with $\mu$ and $\tau$ loops at order
$(\alpha/\pi)^2$ and higher order vacuum polarization and light-by-light
scattering contributions from the muon and the tau at order $(\alpha/\pi)^3$.
In general, these contributions decouple in QED as $(m_e/m_{\mu,\tau})^2$. The
results are known analytically up to three loops. Explicit expressions for
most of the QED contributions can be found in
Ref.~\cite{review_C_M,review_Knecht}, together with the original references.
The error induced by the experimental uncertainty in $m_e/m_{\mu,\tau}$ is
smaller than the digits given in the second and third term. As input values we
used $m_e = 0.510998902(21)~\mbox{MeV}$, $m_\mu = 105.658357(5)~\mbox{MeV}$
and $m_\tau = 1776.99^{+0.29}_{-0.26}~\mbox{MeV}$ from the
PDG~\cite{PDG_2002}, but the independent determination $m_\mu/m_e =
206.768277(24)$ from Ref.~\cite{mmu_over_me}. The four loop result is only
known numerically. Very recently an error in some parts of the contribution
has been found, see Ref.~\cite{Kinoshita_Nio}, which changed the coefficient
of the term $(\alpha/\pi)^4$ by $-0.24$. A numerical evaluation of all terms
is under way~\cite{Kinoshita_Nio} to reduce the error quoted in
Eq.~(\ref{a_e_th}). Certainly, an independent check of this coefficient by
some other group would be highly welcome.

Furthermore, we have included in Eq.~(\ref{a_e_th}) the small hadronic
correction $1.67(3) \times 10^{-12}$ (after correcting the sign in the
light-by-light scattering contribution, see Ref.~\cite{review_Knecht}) and the
electroweak contribution $0.03 \times 10^{-12}$. Since the QED part dominates
over the hadronic and electroweak contribution, one can invert the
relation~(\ref{a_e_th}) in order to get $\alpha$, by comparing $a_e$ with
the experimental value from Eq.~(\ref{a_e_exp}). In this way one obtains
\be \label{alpha_ae} 
\alpha^{-1}(a_e) = 137.035~998~75(50)(13)  
= 137.035~998~75(52) \qquad \mbox{[3.8~ppb]} . 
\ee 
The error found in Ref.~\cite{Kinoshita_Nio} has a very big effect on $a_e$,
shifting it by about $-7.0 \times 10^{-12}$, which changes $\alpha^{-1}$ by
$-8.3 \times 10^{-7}$. This corresponds to $6.1~\mbox{ppb}$, i.e.\ $1.6$
standard deviations. The errors given in Eq.~(\ref{alpha_ae}) come from the
experimental uncertainty in $a_e$ and from the error in the fourth order
coefficient in Eq.~(\ref{a_e_th}), respectively.


\section{Muon anomalous magnetic moment}

The Standard Model contributions are usually split into three parts:
$a_\mu^{\mbox{{\scriptsize SM}}} = a_\mu^{\mbox{{\scriptsize QED}}} +
a_\mu^{\mbox{{\scriptsize EW}}} + a_\mu^{\mbox{{\scriptsize had}}} $. We will
now discuss in turn the three types of contributions.

\subsection{QED contribution}

A general feature here is that electron loops are enhanced due to logarithms
$\ln(m_\mu/m_e) \sim 5.3$. These are short-distance logarithms from vacuum
polarization and infrared logarithms, for instance in the light-by-light
scattering contribution. The latter effect completely
dominates~\cite{review_C_M,review_Knecht} the contribution at order
$(\alpha/\pi)^3$: $a_\mu^{(3)}(\mbox{lbyl}) = \left[ (2/3) \pi^2 \ln (m_\mu /
  m_e) + \ldots \right] \left( \alpha / \pi \right)^3 = 20.947\ldots
\left( \alpha / \pi \right)^3$.  Loops with $\tau$-leptons are again
suppressed. The result in QED up to 5-loops reads
\bea
a_\mu^{\mbox{{\scriptsize QED}}} & = & 0.5 \times \left( {\alpha \over \pi}
\right) + 0.765~857~399(45) \times \left( {\alpha \over \pi} \right)^2
+ 24.050~509~5(23) \times \left( {\alpha \over \pi} \right)^3 \nonumber \\
&&+ 125.08(41) \times \left( {\alpha \over \pi} \right)^4 + 930(170) \times
\left( {\alpha \over \pi} \right)^5 \nonumber \\  
& = & 11~658~470.35(05)(12)(11) \times 10^{-10} 
  = 11~658~470.35 (28) \times 10^{-10}. \label{a_mu_QED} 
\eea
The errors given in the second and third term on the right-hand side are due
to the uncertainty in the experimental values of $m_\mu/m_{e,\tau}$, the one
in the fourth term from the numerical integration and the one in the last term
from a renormalization group estimate of that coefficient~\cite{review_C_M}.
The error found in Ref.~\cite{Kinoshita_Nio} has not such a big effect here.
The coefficient of the term $(\alpha / \pi)^4$ changed by about $-1.0$. This
reduces $a_\mu$ by $-0.29 \times 10^{-10}$. On the other hand, in the first
term one gets a shift of $+0.07 \times 10^{-10}$ due to the change in
$\alpha$. The errors in the last line come from $\alpha$, cf.\ 
Eq.~(\ref{alpha_ae}), from the fourth and from the fifth order coefficient,
respectively. To be on the safe side, we have added them linearly to obtain
the final result, see also Ref.~\cite{review_C_M}. 


\subsection{Hadronic contributions}

\noindent
{\it a) Hadronic vacuum polarization} \\[0.2cm]
We will sketch here only the main issues and give the numerical results of
several recent
evaluations~\cite{Davier_et_al_02,Hagiwara_et_al_02,Jegerlehner_02}. More
details can be found in these references and in Ref.~\cite{Hoecker}.

\begin{figure}[h]
\centerline{\psfig{figure=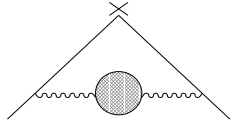,height=1.6cm,width=3cm
}}
\caption{Hadronic vacuum polarization contribution to $a_\mu$.}
\label{fig:a_had_VP} 
\end{figure}

The hadronic corrections induce the largest uncertainties in $a_\mu$, since
they are theoretically not very well under control. The problem is that quarks
are bound by strong gluonic interactions into hadrons at low energies,
relevant for the muon $g-2$. In particular for the light quarks $u,d,s$ one
cannot use perturbative QCD. The coupling $\alpha_s(Q^2)$ is large for $Q^2
\sim 1~\mbox{GeV}^2$ and grows for $Q^2\to 0$. For instance in the hadronic
vacuum polarization contribution $a_\mu^{\mbox{\scriptsize had.\ v.p.}}$,
depicted in Fig.~\ref{fig:a_had_VP}, one cannot simply equate the hadronic
``blob'' with a quark loop as it is possible for a lepton loop.
In the present case there is, however, a way out by using the optical theorem
(unitarity) to relate the imaginary part of the diagram to the measurable
scattering cross section $\sigma(e^+e^- \to \gamma^* \to \mbox{hadrons})$.
From a dispersion relation one then obtains the spectral
representation~\cite{spectral_rep}
\bea
a_\mu^{\mbox{{\scriptsize had.\ v.p.}}} & = & {\alpha \over \pi}
  e^2 \int_0^\infty {ds \over s} {1\over \pi} \mbox{Im} \Pi(s) \int_0^1 dx 
    {x^2 (1-x) \over x^2 + {s \over m_\mu^2} (1-x)} \, ,
    \label{a_had_VP_ImPi} \\ 
{1 \over \pi} \mbox{Im} \Pi(s) & = & {s \over 16 \pi^3 \alpha^2} 
\ \sigma(e^+ e^- \to \gamma^* \to \mbox{hadrons}) \, , \\ 
(q_\mu q_\nu - q^2 \eta_{\mu\nu}) \Pi(q^2) & = &  i \int d^4x e^{i q \cdot x}
\langle \Omega | T \left\{ j_\mu(x) j_\nu(0) \right\} | \Omega \rangle \, . 
\eea
Usually, the relation is expressed as an integral involving the ratio $R(s) =
\sigma(e^+ e^- \to \gamma^* \to \mbox{hadrons}) / \sigma(e^+ e^- \to \gamma^*
\to \mu^+ \mu^-)$ multiplied with a known, positive kernel function peaked at
low-energy.

Information on the spectral function $\mbox{Im} \Pi(s)$ in
Eq.~(\ref{a_had_VP_ImPi}) can also be obtained from hadronic $\tau$ decays,
like $\tau^- \to \nu_\tau \pi^- \pi^0$. One has, however, to apply corrections
due to isospin violations, since $m_u \neq m_d$ and because of electromagnetic
radiative corrections, see
Refs.~\cite{Cirigliano_et_al,Davier_et_al_02,Hoecker}. 

The most recent estimates are collected in Table~\ref{tab:a_mu_had_VP}. 
%
\begin{table}[h] 
\caption{Recent evaluations of $a_{{\scriptscriptstyle
\mu}}^{\mbox{{\tiny had.\ v.p.}}}$} 
\label{tab:a_mu_had_VP}   
\begin{center}
\renewcommand{\arraystretch}{1.1}
\begin{tabular}{|l|r@{.}lcr@{.}l|}
\hline 
Authors & \multicolumn{5}{|c|}{Contribution to
$a_\mu \times 10^{10}$}  \\ 
\hline 
Davier et al.~\cite{Davier_et_al_02} ($e^+ e^- + \tau$)   & 709 & 0 & $\pm$ &
5 & 9  \\ 
Davier et al.~\cite{Davier_et_al_02} ($e^+ e^-$)  & \hspace*{1cm} 684 & 7 &
$\pm$ & 7 & 0 \\ 
Hagiwara et al.~\cite{Hagiwara_et_al_02} ($e^+ e^-$)      & 683 & 1 & $\pm$ & 6
& 2  \\ 
Jegerlehner~\cite{Jegerlehner_02} ($e^+ e^-$)           & 683 & 62 & $\pm$ & 8
& 61 \\ 
\hline 
\end{tabular}
\end{center} 
\end{table} 
One observes that the evaluations based on $e^+e^-$ and $\tau$ data in
Ref.~\cite{Davier_et_al_02} are inconsistent with each other at the
$2.5~\sigma$ level. Note~\cite{Hoecker_private} that at least part of this
discrepancy could be due to an error in the cross section measured by the
CMD-2 collaboration~\cite{CMD_2}. This measurement with its small uncertainty
dominates at present the low-energy region around the $\rho$-peak and
therefore the final result for $a_\mu^{\mbox{\scriptsize had.\ v.p.}}$.  After
correcting the error, the value will then shift towards the one based on
$\tau$ data. Still, there remain quite large discrepancies between the
spectral functions derived from $e^+e^-$ and $\tau$ data in some energy
regions above the $\rho$ which are not yet understood and which cannot be
explained by the known sources of isospin violation. Presumably, only future
measurements of $\sigma(e^+ e^- \to \mbox{hadrons})$, either using the
radiative return method~\cite{radiative_return} at KLOE (Daphne) or BABAR or
with a new scan at VEPP-2000, will be able to resolve the puzzle. In any case,
it will be necessary to better understand the implementation of radiative
corrections to the hadronic final state~\cite{sigma_had_rad_corr}.

Averaging the results that use $e^+e^-$ data only, we obtain 
\be \label{a_mu_had_VP_ee}
a_\mu^{\mbox{{\scriptsize had.\ v.p.}}}(e^+e^-) = 
683.8(7.5) \times 10^{-10} \, .  
\ee

Recently, a first evaluation of $a_\mu^{\mbox{{\scriptsize had.\  v.p.}}}$ on
the lattice appeared, although still with very large uncertainties~\cite{Blum}
\be
\left. a_\mu^{\mbox{{\scriptsize had.\ v.p.}}} \right|_{u,d,s} =
  460(78) \times 10^{-10} \, . 
\ee 
Note that the error is only statistical. Large systematical errors from
the quenching approximation, unphysically large quark masses and finite volume
effects are not accounted for. It will probably take a very long time to even
come down to a 10\% error. 

Finally, there are higher order vacuum polarization effects, if additional
photonic corrections or fermion loops (leptons and hadrons) are added to the
diagram in Fig.~\ref{fig:a_had_VP}. They have been evaluated in
Ref.~\cite{Krause} with the result 
\be \label{a_mu_had_ho_VP}
a_\mu^{\mbox{{\scriptsize h.o.-h.v.p.}}} = - 10.0(0.6)
\times 10^{-10} \, . 
\ee

\begin{figure}[h]
\epsfxsize=30pc 
\centerline{\epsfbox{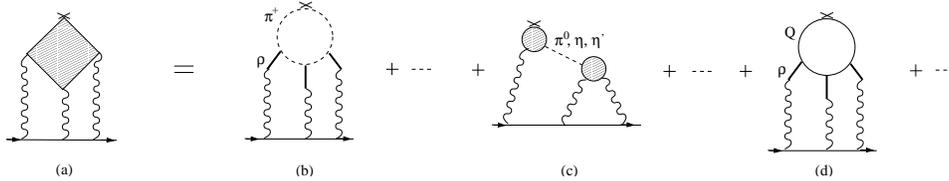}} 
\caption{The hadronic light-by-light scattering contribution to the
muon $g-2$.}
\label{fig:overview}
\end{figure}

\vspace*{0.1cm} 
\noindent
{\it b) Hadronic light-by-light scattering} \\[0.2cm]
The present picture of hadronic light-by-light scattering, as reviewed
recently in Ref.~\cite{LbyL_reviews}, is shown in Fig.~\ref{fig:overview} 
and the corresponding contributions to $a_\mu$ are listed in
Table~\ref{tab:overview}, taking into account the 
corrections made in the two full evaluations~\cite{HKS_corr,BPP_corr},
after we had discovered the sign error in the pion-pole
contribution~\cite{KN_pion,a_mu_EFT}. 
%
\begin{table}[h]
\caption{Contributions to $a_{{\scriptscriptstyle
\mu}} (\times 10^{10})$ according to Fig.~\ref{fig:overview}. } 
\label{tab:overview}   
\begin{center}
\renewcommand{\arraystretch}{1.1}
\begin{tabular}{|l|r@{.}l|r@{.}l|c|c|}
\hline
Type &
\multicolumn{2}{|c|}{{Ref.~\cite{HKS_corr}}} & 
\multicolumn{2}{|c|}{{Ref.~\cite{BPP_corr}}} & 
Ref.~\cite{KN_pion} & No form factors \\ 
\hline  
(b)                & -0 & 5(0.8) & -1 & 9(1.3) & & -4.5 \\ 
(c) & 8 & 3(0.6) & 8 & 5(1.3) & 8.3(1.2)    & $+\infty$ \\
$f_0, a_1$               & 0 & 174$^{\rm a}$ & -0 & 4(0.3) &   
&  \\
(d) & 1 & 0(1.1) & 2 & 1(0.3) &    & $\sim 6$ \\
\hline
Total & 9 & 0(1.5) & 8 & 3(3.2) & 8(4)$^{\rm b}$ &  \\ 
\hline 
\multicolumn{7}{l}{{\footnotesize $^{\rm a}$~Only $a_1$ exchange.}} \\
\multicolumn{7}{l}{{\footnotesize $^{\rm b}$~Our estimate, using
  Refs.~\cite{HKS_corr,BPP_corr,KN_pion}.}}  
\end{tabular}
\end{center}
\end{table}

There are three classes of contributions to the hadronic four-point function
[Fig.~\ref{fig:overview}(a)], which can be understood from an effective field
theory (EFT) analysis of hadronic light-by-light
scattering~\cite{EdeR_94,a_mu_EFT}: (1) a charged pion loop
[Fig.~\ref{fig:overview}(b)], where the coupling to photons is dressed by some
form factor ($\rho$-meson exchange, e.g.\ via vector meson dominance (VMD)),
(2) the pseudoscalar pole diagrams [Fig.~\ref{fig:overview}(c)] together with
the exchange of heavier resonances ($f_0, a_1, \ldots$) and, finally, (3) the
irreducible part of the four-point function which was modeled in
Refs.~\cite{HKS_corr,BPP_corr} by a constituent quark loop dressed again with
VMD form factors [Fig.~\ref{fig:overview}(d)]. The latter can be viewed as a
local contribution $\bar\psi \sigma^{\mu\nu} \psi F_{\mu\nu}$ to $a_\mu$. The
two groups~\cite{HKS_corr,BPP_corr} used similar, but not identical models
which explains the slightly different results for the dressed charged pion and
the dressed constituent quark loop, although their sum seems to cancel to a
large extent and the final result is essentially given by the pseudoscalar
exchange diagrams. We take the difference of the results as indication of the
error due to the model dependence.

\vspace*{0.2cm} 
\noindent
{\it Pion-pole contribution} \\[0.1cm]
The contribution from the neutral pion intermediate state is given by a
two-loop integral that involves the convolution of two pion-photon-photon
transition form factors $\FF$, see Fig.~\ref{fig:overview}(c). We refer to
Ref.~\cite{KN_pion} for all the details. Since no data on the doubly off-shell
form factor $\FF(q_1^2,q_2^2)$ is available, one has to resort to models. In
order to proceed with the analytical~\footnote{In contrast, the calculations
  in Refs.~\cite{HKS_corr,BPP_corr} were based purely on numerical
  approaches.} evaluation of the two-loop integral, we considered a certain
class of form factors which includes the ones based on large-$N_C$ QCD that we
studied in Ref.~\cite{KN_VAP}.  For comparison, we have also used a vector
meson dominance (VMD) and a constant form factor, derived from the
Wess-Zumino-Witten (WZW) term.

Our approach to this problem consists of making an ansatz for the relevant
Green's functions in the framework of large-$N_C$ QCD. In this limit, an
infinite set of narrow resonance states contributes in each channel. The
pion-photon-photon form factor is then given by a sum over an infinite set of
narrow vector resonances, involving arbitrary couplings, although there are
constraints at long and short distances. To implement them, we perform a
matching of the ansatz with chiral perturbation theory (ChPT) at low energies
and with the operator product expansion (OPE) at high momenta in order to
reduce the model dependence. In practice, it is sufficient to keep a few
resonance states to reproduce the leading behavior in ChPT and the OPE.  The
normalization is given by the WZW term, $\FF(0,0) = - N_C / (12 \pi^2 F_\pi)$,
whereas the OPE tells us that $\lim_{\lambda\to \infty}\,\FF(\lambda^2 q^2,
(p-\lambda q)^2) = {(2 F_\pi) / (3 \lambda^2 q^2)} + \order(1/\lambda^3)$. We
considered the form factors that are obtained by truncation of the infinite
sum to one (lowest meson dominance, LMD), and two (LMD+V), vector resonances
per channel, respectively. Some of the parameters in the LMD+V form factor are
not fixed by the normalization and the leading term in the OPE. We have
determined these coefficients phenomenologically~\cite{KN_VAP,KN_pion}.  In
particular, $\FF(\!-Q^2,\!0)$ with one photon on-shell behaves like $1/Q^2$
for large spacelike momenta, $Q^2\!=\!-q^2$. Whereas the LMD form factor does
not have such a behavior, it can be reproduced with the LMD+V ansatz.  Note
that the usual VMD form factor $\FF^{\mbox{\tiny VMD}}(q_1^2,q_2^2) \sim 1/
[(q_1^2 - M_V^2) (q_2^2 - M_V^2)]$ does {\it not} correctly reproduce the OPE.

For the form factors discussed above one can perform {\it all} angular
integrations in the two-loop integral analytically~\cite{KN_pion}.  The
pion-exchange contribution to $a_\mu$ can then be written as a two-dimensional
integral representation, where the integration runs over the moduli of the
Euclidean momenta
\be
a_{\mu}^{\mbox{\tiny{LbyL;$\pi^0$}}} = \int_0^\infty d Q_1 
\int_0^\infty d Q_2 \ \sum_i w_i(Q_1,Q_2) \ f_i(Q_1,Q_2) ,  
\ee
with universal [for the above class of form factors] weight functions $w_i$
(rational functions, square roots and logarithms)~\cite{KN_pion}. The
dependence on the form factors resides in $f_i$. In this way we could separate
the generic features of the pion-pole contribution from the model dependence
and thus better control the latter. This is not possible anymore in the final
analytical result (as a series expansion) for
$a_{\mu}^{\mbox{\tiny{LbyL;$\pi^0$}}}$ derived in Ref.~\cite{Blokland_etal}.
Note that the analytical result has not the same status here as for instance
in QED. One has to keep in mind that there is an intrinsic uncertainty in the
form factor of $10 - 30\%$, furthermore the VMD form factor used in that
reference has the wrong high-energy behavior.

The weight functions $w_i$ in the main contribution are positive and peaked
around momenta of the order of $0.5~\mbox{GeV}$. There is, however, a tail in
one of these functions, which produces for the constant WZW form factor a
divergence of the form $\ln^2\!\Lambda$ for some UV-cutoff $\Lambda$. We will
come back to this point below. Other
weight functions have positive and negative contributions in the low-energy
region, which lead to a strong cancellation in the corresponding integrals.

All form factors lead to very similar results (apart from WZW). Judging from
the shape of the weight functions described above, it seems more important to
correctly reproduce the slope of the form factor at the origin and the
available data at intermediate energies. On the other hand, the asymptotic
behavior at large $Q_i$ seems not very relevant.  The results for the LMD+V
form factor are rather stable under the variation of the parameters.

With the LMD+V form factor, we then get 
\be 
a_{\mu}^{\mbox{\tiny{LbyL;$\pi^0$}}} = + 5.8(1.0) \times 10^{-10}
\, , 
\ee
where the error includes the variation of the parameters and the intrinsic
model dependence. A similar short-distance analysis in the framework of
large-$N_C$ QCD and including quark mass corrections for the form factors for
the $\eta$ and $\eta^\prime$ was beyond the scope of Ref.~\cite{KN_pion}. We
therefore used VMD form factors fitted to the available data for $\FF(-Q^2,0)$
to obtain our final estimate
\be
a_{\mu}^{\mbox{\tiny{LbyL;PS}}} \equiv  
a_{\mu}^{\mbox{\tiny{LbyL;$\pi^0$}}}
+ a_{\mu}^{\mbox{\tiny{LbyL;$\eta$}}}\vert_{\mbox{\tiny VMD}} +
a_{\mu}^{\mbox{\tiny{LbyL;$\eta^\prime$}}}\vert_{\mbox{\tiny VMD}} 
=  + 8.3(1.2) \times 10^{-10} \, . 
\ee
An error of 15~\% for the pseudoscalar pole contribution seems reasonable,
since we impose many theoretical constraints from long and short distances on
the form factors. Furthermore, we use experimental information whenever
available.


\vspace*{0.2cm} 
\noindent
{\it Effective field theory approach to $a_{\mu}^{\mbox{\tiny{LbyL;had}}}$}
\\[0.1cm] 
In Ref.~\cite{a_mu_EFT} we discussed an EFT approach to hadronic
light-by-light scattering based on an effective Lagrangian that describes the
physics of the Standard Model well below 1~GeV, see also Ref.~\cite{EdeR_94}.
It includes photons, light leptons, and the pseudoscalar mesons and obeys
chiral symmetry and $U(1)$ gauge invariance.

The leading contribution to $a_{\mu}^{\mbox{\tiny{LbyL;had}}}$, of order
$p^6$, is given by a finite loop of charged pions with point-like
electromagnetic vertices, see Fig.~\ref{fig:overview}(b).  Since this
contribution involves a loop of hadrons, it is subleading in the large-$N_C$
expansion.

\begin{figure}[h]
\epsfxsize=18pc 
\centerline{\epsfbox{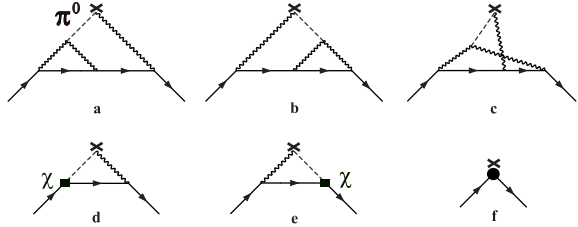}} 
\caption{The graphs contributing to
$a_{\mu}^{\mbox{\tiny{LbyL;$\pi^0$}}}$ at lowest order in the effective
field theory.} 
\label{fig:EFT}
\end{figure}

At order $p^8$ and at leading order in $N_C$, we encounter the divergent
pion-pole contribution, diagrams (a) and (b) of Fig.~\ref{fig:EFT}, involving
two WZW vertices.  The diagram (c) is actually finite. The divergences of the
triangular subgraphs in the diagrams (a) and (b) are removed by inserting the
counterterm $\chi$ from the Lagrangian $\lag^{(6)} = (\alpha^2 / 4 \pi^2 F_0)
\ \chi \ {\overline\psi} \gamma_\mu \gamma_5 \psi \, \partial^\mu \pi^0 +
\cdots$, see the one-loop diagrams (d) and (e). Finally, there is an overall
divergence of the two-loop diagrams (a) and (b) that is removed by a local
counterterm, diagram (f).  Since the EFT involves such a local contribution,
we will not be able to give a precise numerical prediction for
$a_{\mu}^{\mbox{\tiny{LbyL;had}}}$.

Nevertheless, it is interesting to consider the leading and next-to-leading
logarithms that are in addition enhanced by a factor $N_C$ and which can be
calculated using the renormalization group~\cite{a_mu_EFT}. The EFT and
large-$N_C$ analysis tells us that
\bea
a_{\mu}^{\mbox{\tiny{LbyL;had}}} & = & 
\left( {\alpha \over \pi} \right)^3  \Bigg\{
f\left({M_{\pi^\pm} \over m_\mu}, {M_{K^\pm} \over m_\mu}\right)
 + N_C \left( {m_\mu^2 \over 16 \pi^2
F_\pi^2} {N_C \over 3} \right)
\left[ \ln^2 {\mu_0 \over m_\mu} + c_1 \ln {\mu_0 \over m_\mu} + c_0
\right]  \nonumber \\ 
&&\qquad\quad + \order \left(\!{m_\mu^2 \over \mu_0^2} \times
\mbox{log's}\!\right) + \order \left(\!{m_\mu^4 \over \mu_0^4} 
N_C \times \mbox{log's}\!\right)\!\!\Bigg\},  \label{a_mu_EFT_N_C}
\eea
where $f(M_{\pi^\pm}\!/\!m_\mu,\!M_{K^\pm}\!/\!m_\mu)\!=\!\!-0.038$ represents
the charged pion and kaon-loop that is formally of order one in the chiral and
$N_C$ counting and $\mu_0$ denotes some hadronic scale, e.g.\ $M_\rho$.  The
coefficient ${\cal C} = (N_C^2 m_\mu^2) / (48 \pi^2 F_\pi^2) = 0.025$ of the
log-square term in the second line is universal and of order $N_C$, since
$F_\pi\!=\!\order(\sqrt{N_C})$. The value given corresponds to $N_C = 3$.

Unfortunately, although the logarithm is sizeable, $\ln(M_\rho/m_\mu) = 1.98$,
in $a_{\mu}^{\mbox{\tiny{LbyL;$\pi^0$}}}$ there occurs a cancellation between
the log-square and the log-term. If we fit our result for the VMD form factor
for large $M_\rho$ to an expression as given in Eq.~(\ref{a_mu_EFT_N_C}), we
obtain
\bea
a_{\scriptscriptstyle{\mu;\mbox{\tiny{VMD}}}}^{\mbox{\tiny{LbyL}}; \pi^0}
& \doteq &\!\!\left( {\alpha \over \pi} \right)^3 {\cal C} 
\ \ \left[ \ln^2 {M_\rho \over m_\mu} + c_1 \ln {M_\rho \over m_\mu} + c_0
\right] \stackrel{\mbox{\tiny{Fit}}}{=} \left( {\alpha \over  \pi}
\right)^3 {\cal C}  
\ \ \left[ 3.94 - 3.30 + 1.08 \right] \nonumber \\
& = & \left[ 12.3 - 10.3 + 3.4 \right] \times 10^{-10}
= 5.4 \times 10^{-10} \, , 
\eea
which is confirmed by the analytical result of Ref.~\cite{Blokland_etal}
(setting for simplicity $M_{\pi^0} = m_\mu$): $a_{\scriptscriptstyle{\mu;
    \mbox{\tiny{VMD}}}}^{\mbox{\tiny{LbyL}}; \pi^0} = [12 - 8.0 + 1.7] \times
10^{-10} = 5.7 \times 10^{-10}$. This cancellation is now also visible in the
published version of Ref.~\cite{Ramsey-Musolf_Wise}. In that paper the
remaining parts of $c_1$ have been calculated: $c_1 = - 2 \chi(\mu_0) / 3 +
0.237 = -0.93^{+0.67}_{-0.83}$, with our conventions for $\chi$ and
$\chi(M_\rho)_{{\rm exp}} = 1.75^{+1.25}_{-1.00}$.

Finally, the EFT analysis shows that the modeling of hadronic light-by-light
scattering by a constituent quark loop is not consistent with QCD.  The latter
has a priori nothing to do with the full quark loop in QCD which is dual to
the corresponding contribution in terms of hadronic degrees of freedom.
Equation~(\ref{a_mu_EFT_N_C}) tells us that at leading order in $N_C$ any
model of QCD has to show the behavior $a_\mu^{\mbox{\tiny{LbyL;had}}} \sim
(\alpha/\pi)^3 N_C [N_C m_\mu^2 / (48 \pi^2 F_\pi^2)] \ln^2\Lambda$, with a
universal coefficient ${\cal C}$, if one sends the cutoff $\Lambda$ to
infinity. From the analytical result for the quark loop, one obtains the
behavior $a_\mu^{\mbox{\tiny LbyL;CQM}} \sim (\alpha/\pi)^3 N_C (m_\mu^2 /
M_Q^2) + \ldots$, for $M_Q \gg m_\mu$, if we interpret the constituent quark
mass $M_Q$ as a hadronic cutoff.  Even though one may argue that $N_C / (48
\pi^2 F_\pi^2)$ can be replaced by $1/M_Q^2$, the log-square term is not
correctly reproduced with this model. Therefore, the constituent quark model
(CQM) cannot serve as a reliable description for the dominant contribution to
$a_\mu^{\mbox{\tiny{LbyL;had}}}$, in particular, its sign.  Moreover, we note
that the pion-pole contribution is {\it infrared finite} in the chiral
limit,\footnote{This can be shown by studying the low momentum behavior of the
  weight functions $w_i$ corresponding to the two-loop \nopagebreak[1]
  diagrams 3(a)--(c) and the one-loop diagrams 3(d)+(e) (given in
  Ref.~\cite{a_mu_EFT}) for $M_{\pi^0} \to 0$, see also
  Ref.~\cite{Ramsey-Musolf_Wise}.}  whereas the quark loop shows an infrared
divergence $\ln(M_Q/m_\mu)$ for $M_Q \to 0$.

The analysis within the EFT and large-$N_C$ framework, together with the
numerical results for all contributions depicted in Fig.~\ref{fig:overview}
and listed in Table~\ref{tab:overview} leads us to the following
(conservative) estimate for the hadronic light-by-light scattering
contribution
\be \label{a_mu_LbyL_had}
a_{\mu}^{\mbox{\tiny{LbyL;had}}} = +\,8(4) \times 10^{-10} \, . 
\ee
Since the model calculations for the dressed charged pion and the dressed
constituent quark loop yield slightly different results we have added the
errors linearly. 


\vspace*{-1mm} 
\subsection{Electroweak contribution}
 
The electroweak correction to $a_\mu$ lies somehow in between the
QED and hadronic contribution. At one loop, the result is reliably
calculable~\cite{EW_1loop}  
\be
a_\mu^{\mbox{\scriptsize EW, (1)}} = 19.5 \times 10^{-10} \, ,
\label{a_mu_EW_1loop} 
\ee
with a Higgs boson contribution that is very small for $M_H \geq
114.5~\mbox{GeV}$ (LEP 2 bound).  

\begin{figure}[h]
\centerline{\psfig{figure=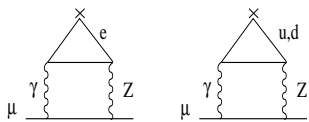,height=1.5cm,width=4cm
}}
\caption{Two-loop electroweak corrections to $a_\mu$ from the first
  family. The light quark loop is to be understood \newline 
\hspace*{-4.75cm} symbolically, representing again a hadronic ``blob'' (QCD
three-point function).}  
\label{fig:aEW_2loops} 
\end{figure}

Two-loop corrections, see Fig.~\ref{fig:aEW_2loops}, 
are potentially large due to factors $\ln(M_Z/m_\mu) \sim 6.8$.  Furthermore,
as noted in Ref.~\cite{Peris_et_al}, one cannot separate leptons and quarks
anymore, but must treat each generation together because of the cancellation
of the triangle anomalies.  Therefore earlier estimates~\cite{Kukhto_et_al}
were incomplete. A first full two-loop calculation was done in
Ref.~\cite{EW_2loops_CKM} and recently revisited by two
groups~\cite{EW_2loops_Knecht_et_al,EW_2loops_CMV} to improve on the treatment
of the hadronic contributions. Instead of using a simple constituent quark
loop, short-distance constraints from the OPE on the relevant QCD three-point
functions have been imposed. There is still some disagreement in the details,
but the numerical values are very close. Adding the one-loop result from
Eq.~(\ref{a_mu_EW_1loop}), Ref.~\cite{EW_2loops_Knecht_et_al} obtains
$a_\mu^{\mbox{\tiny EW}} = 15.2(0.1) \times 10^{-10}$. The error reflects
hadronic uncertainties and the variation of $M_H$. No
resummation~\footnote{The resummation of leading logarithms has been discussed
  in Ref.~\cite{Degrassi_Giudice} and corrected in Ref.~\cite{EW_2loops_CMV}.}
has been performed in that reference.  Ref.~\cite{EW_2loops_CMV} gets
$a_\mu^{\mbox{\tiny EW}} = 15.4(0.1)(0.2) \times 10^{-10}$, where the first
error corresponds to the hadronic uncertainty and the second to an allowed
Higgs boson mass range of $114~\mbox{GeV} \leq M_H \leq 250~\mbox{GeV}$, the
current top mass uncertainty, and unknown three-loop effects. There are large
cancellation in the resummation, therefore the final shift after the
resummation is very small. Averaging the two estimates, we obtain
\be \label{a_mu_EW}
a_\mu^{\mbox{\scriptsize EW}} = 15.3(0.2) \times 10^{-10} \, , 
\ee
which corresponds to quite a large two-loop correction of
$a_\mu^{\mbox{\tiny EW, (2)}} = - 4.2(0.2) \times 10^{-10}$.
Although not all details have been resolved, the electroweak contribution
seems well under control.


\subsection{Summary} 

We now collect the results for the different contributions in the Standard
Model from Eqs.~(\ref{a_mu_QED}), Table~\ref{tab:a_mu_had_VP},
(\ref{a_mu_had_VP_ee}), (\ref{a_mu_had_ho_VP}), (\ref{a_mu_LbyL_had}), and
(\ref{a_mu_EW}):
\bea
a_\mu^{\mbox{{\scriptsize SM}}}(e^+ e^-) & = &
(11~659~167.5 \pm 7.5 \pm 4.0 \pm 0.35) \times  10^{-10} \, , \\
a_\mu^{\mbox{{\scriptsize SM}}}(\tau)  & = &
(11~659~192.7 \pm \underbrace{5.9}_{\mbox{\scriptsize v.p.}} \pm
  \underbrace{4.0}_{\mbox{\scriptsize LbyL}} \pm
  \underbrace{0.35}_{\mbox{\scriptsize EW} }) \times  10^{-10}  
  \, ,  
\eea
where we kept the results based on $e^+e^-$ and $\tau$ data separately.   
Comparison with the experimental value from Eq.~(\ref{a_mu_exp}) leads to 
\bea
a_\mu^{\mbox{{\scriptsize exp}}} - a_\mu^{\mbox{{\scriptsize SM}}}(e^+ e^-) 
& = & (35.5 \pm 11.7) \times 10^{-10} \hspace*{1.275cm} [3.0~\sigma] \, , \\  
a_\mu^{\mbox{{\scriptsize exp}}} - a_\mu^{\mbox{{\scriptsize SM}}}(\tau) 
& = & (10.3 \pm 10.7) \times 10^{-10} \hspace*{1.275cm} [1.0~\sigma] \, .  
\eea
Is the discrepancy using the $e^+e^-$ data a sign for new physics beyond the
Standard Model ? In view of the inconsistencies between the evaluations based
on $e^+e^-$ and $\tau$ data this conclusion is certainly premature.
Furthermore, the error found in the CMD-2 data~\cite{Hoecker_private} will
probably reduce the discrepancy between $a_\mu^{\mbox{{\scriptsize exp}}}$ and
$a_\mu^{\mbox{{\scriptsize SM}}}(e^+ e^-)$ to less than 2~$\sigma$.


\section{Conclusions} 

We briefly presented the current theoretical value for the anomalous magnetic
moment of the electron. An error found recently in the coefficient of the
four-loop QED result leads to a $1.6~\sigma$ shift in the fine structure
constant $\alpha$ when comparing theoretical and experimental values
for~$a_e$.

We then reviewed the prediction for the anomalous magnetic moment of the muon
$a_\mu$ in the Standard Model. In particular, we discussed the uncertainties
induced by the hadronic contributions, like vacuum polarization,
light-by-light scattering and higher order electroweak corrections. Using
$e^+e^-$ data, an apparent discrepancy between Standard Model value and
experimental value exists which could point to new physics. In view of
inconsistencies when using $e^+e^-$ and $\tau$ data and because there seems to
be an error in the latest, most precise $e^+e^-$ data, such a conclusion
cannot be drawn yet.  Perhaps some of the hadronic uncertainties are even
underestimated. There still remains a lot of work to be done to better control
these hadronic contributions, if we want to use the muon $g-2$ to search for
signs of new physics.


\section*{Acknowledgments}

\vspace*{-0.2cm}
I am grateful to my collaborators Marc Knecht, Michel Perrottet and Eduardo de
Rafael. Furthermore, I thank Vincenzo Cirigliano, Fred Gray, Andreas H\"ocker,
Axel Hoefer, Fred Jegerlehner and Ximo Prades for discussions. I would also
like to thank the organizers of the Rencontres de Moriond for the invitation
and for providing such a pleasant atmosphere.


\end{document}